# Motion Compensated Extreme MRI: Multi-Scale Low Rank Reconstructions for Highly Accelerated 3D Dynamic Acquisitions (MoCo-MSLR)

Zachary Miller[1*], Luis Torres[2*], Sean Fain[4], Kevin Johnson[2,3]

*co first-authors

1. Department of Biomedical Engineering, University of Wisconsin, Madison, Wisconsin, USA

2. Department of Medical Physics, University of Wisconsin School of Medicine and Public Health, Madison, Wisconsin, USA

3. Department of Radiology, University of Wisconsin School of Medicine and Public Health, Madison, Wisconsin, USA

4. Department of Radiology, Carver College of Medicine, University of Iowa Health Care, Iowa City, Iowa, USA

**Abstract:**
Purpose: To improve upon Extreme MRI, a recently proposed method by Ong Et al. for reconstructing high spatiotemporal resolution, 3D non-Cartesian acquisitions by incorporating motion compensation into these reconstructions using an approach termed MoCo-MSLR.
Methods: Motion compensation is challenging to incorporate into high spatiotemporal resolution reconstruction due to the memory footprint of the motion fields and the potential to lose dynamics by relying on an initial high temporal resolution, low spatial resolution reconstruction. Motivated by the work of Ong Et al. and Huttinga Et al., we estimate low spatial resolution motion fields through a loss enforced in k-space and represent these motion fields in a memory efficient manner using multi-scale low rank components. We interpolate these motion fields to the desired spatial resolution, and then incorporate these fields into Extreme MRI.
Results: MoCo-MSLR was able to improve image quality for reconstructions around 500ms temporal resolution and capture bulk motion not seen in Extreme MRI. Further, MoCo-MSLR was able to resolve realistic cardiac dynamics at near 100ms temporal resolution while Extreme MRI struggled to resolve these dynamics.
Conclusion: MoCo-MSLR improved image quality over Extreme MRI and was able to resolve both respiratory and cardiac motion in 3D.

## 1. Introduction

In recent years, significant work has gone towards development of free-breathing, high spatiotemporal resolution 4D acquisitions (1),(2). These acquisitions combined with robust reconstruction methods have the potential to reduce the challenge of imaging pediatric (3) and neonatal subjects and allow patients with severe cardiopulmonary disease to feel more comfortable during scanning by removing the need for breath-holds (4). These methods can also provide improved spatiotemporal resolution for dynamic contrast-enhanced acquisitions with implications for the visualization and quantification of functional measures of hemodynamics and contrast uptake. In addition, these methods provide significant advantages for thoracic imaging, where motion corruption is common and difficult to manage (5) .

These dynamic acquisitions are often acquired using non-Cartesian methods with pseudorandom view ordering. One of the benefits of this approach is that acquired data can be flexibly re-binned after the acquisition. This allows reconstructions across multiple dimensions in order to, for instance, resolve respiratory and cardiac motion. These binning methods are often performed using surrogate motion signals derived using respiratory belts, pilot tone modulation, or center of k-space based navigators (6) (7) (8). Using the motion surrogate, k-space data is typically binned prior to image reconstruction into a small number of motion states with the assumption these motion states recur periodically through the acquisition. In acquisitions with irregular respiratory or bulk motion, however, reconstruction performance using these binning techniques can be significantly degraded due to artifact from intraframe motion (6).

One approach to solving this problem is to bin data through time with sufficient temporal resolution (e.g. for respiratory motion ~500ms) to reduce intraframe motion. Reconstructing such data, however, is challenging due to the extreme degree of undersampling of individual frames and sheer amount of data generated by binning at sub-second intervals in minutes long scans. For smaller scale problems (e.g. lower spatiotemporal resolution), techniques that leverage correlations across frames via nuclear norm minimization are often used to reconstruct highly undersampled data (9). However, with increased matrix size and frame count, nuclear norm minimization quickly become infeasible with respect to memory and computation time (6).

Ong et al (6) proposed a way to overcome this memory and computational bottleneck by directly optimizing for a highly compressed multi-scale low rank (MSLR) representation of the 4D time series. This method dubbed "Extreme MRI", is not only able to capture irregular and bulk motion in free breathing high spatiotemporal ultrashort echo time (UTE) pulmonary and DCE MRI acquisitions (3), but is able to further reduce the rank of the data set directly in the compressed space using the variational definition of the nuclear norm (10,11).

Like all low rank methods though, Extreme MRI is dependent on correlations across frames. Bulk and irregular motion disrupts these correlations and erodes image quality. We hypothesized that incorporating motion compensation into Extreme MRI would improve image quality as it improves these correlations. This hypothesis is supported by a large body of work showing that incorporating motion compensation into reconstruction significantly improves reconstruction quality (5,12)

Much of this work, however, relies on motion field estimation through retrospective registration of low-resolution navigator images. This is problematic if the initial low-resolution reconstruction is unable to capture all motion dynamics. In the case of Extreme MRI at high temporal resolution (<500ms per frame), the accurate reconstruction of low resolution images themselves is challenging due to high levels of undersampling. Furthermore, many of these motion correction algorithms operate on relatively small-scale problems where memory constraints are less of a concern. For the scale of the problems Extreme MRI is attempting to reconstruct, use of dense motion fields can easily triple the memory footprint of reconstruction.

In recent work, Huttinga et al. (13) have overcome these constraints by developing memory efficient methods to estimate motion fields directly from k-space data binned through time. This method warps a reference image-template according to loss enforced in k-space, and directly solves for a cubic B-spline parameterization of low rank representations of the motion fields. Using this method, Huttinga et al. can recover respiratory motion up to 100ms temporal resolution. As they use a k-space representation of motion fields relative to one static frame, they do not need prior dynamic reconstructions to accurately model motion.

In this work, motivated by the developments in (13) and (6), we integrate a memory efficient representation of the motion fields estimated in k-space with Extreme MRI reconstructions. Our proposed method involves estimating low resolution motion fields directly as multi-scale low rank components by enforcing k-space loss between a warped template image and acquired k-space data, up-sampling these fields directly in the compressed space, updating these fields through k-space based loss when needed, and then integrating these high-resolution fields into Extreme MRI. We apply this Motion Corrected MSLR technique (MoCo-MSLR) to 3D free breathing radial acquisitions and compare it to Extreme MRI reconstructions at temporal resolutions required to resolve respiratory (500ms) and cardiac dynamics (100ms).

## 2. Theory
### 2.1 Extreme MRI: Multi-scale Low Rank Reconstruction Review

The MSLR model (6,14) stacks a time series with $T$ frames and image size $N$ into a spatiotemporal matrix $X$ of size $T \times N$. This spatiotemporal matrix is then represented as the sum of rank 1 block-wise matrices across varying block size scales. If $J$ is the number of block scales for the MSLR decomposition then for a given block scale $j \in J$, $Bl_j$ blocks of size $N_j \times T$ are returned which are then factored into a block-wise left spatial basis $L_j \in \mathbb{C}^{N_j \times 1}$ and a right temporal basis $R_j \in \mathbb{C}^{T \times 1}$. The sum of this decomposition across block-sizes for a frame $X_t$ is:

$$X_t = \sum_{j=1}^{J} \boldsymbol{B_j}(L_j R_{j,t}^H) \ (1)$$

Where $\boldsymbol{B_j}$ is a block-to-image operator.

The forward model for the reconstruction problem then with acquired multi-channel k-space data stacked into a matrix $Y \in \mathbb{C}^{CM \times T}$ where C is the number of coils, $M$ is the number of measurements and $T$ is number of frames is:

$$Y_t = \mathcal{A}\left(\sum_{j=1}^{J} B_j\left(L_j R_{j,t}^H\right)\right) \quad (2)$$

Where $\mathcal{A}$ is a linear operator incorporating sensitivity maps and the non-uniform fast Fourier transform operator. To regularize the problem, Ong et al. applies block-wise low rank constraints by using the variational form of nuclear norm minimization:

$$\min_{X=\sum_{j=1}^{J} M_j(L_j R_j)} \|X\|_* = \sum_{j=1}^{J}(\|L_j\|_F^2 + \|R_j\|_F^2) \quad (3)$$

This formulation allows for block-wise rank reduction directly in the compressed space significantly reducing the memory and computational requirements associated with computing the nuclear norm. The full MSLR reconstruction objective function to be minimized is:

$$f(L,R) = \frac{1}{2}\left\|Y - \mathcal{A}\left(\sum_{j=1}^{J} B_j\left(L_j R_j^H\right)\right)\right\|^2 + \frac{\lambda_j}{2}\sum_{j=1}^{J}(\|L_j\|_F^2 + \|R_j\|_F^2) \quad (4)$$

To further reduce reconstruction run-time , stochastic optimization is used to solve for the right and left vectors, taking gradient steps frame by frame rather than averaging across all frames.

In the formulation in **(4)**, the MSLR factorization attempts to model all dynamics including motion and contrast change in the time series. The greater the complexity of dynamics contained in this decomposition, the higher the rank must be to appropriately model these dynamics. As the decomposition intrinsically constrains rank, complex dynamics that cannot be modeled in this setting can be lost resulting in artifacts, blurring, and/or misrepresentation of the dynamics. Irregular respiratory and bulk motion is particularly challenging to model as it is usually associated with high rank.

### 2.2 MoCo-MSLR Reconstruction

Let **forward** motion fields be defined as warps from a fixed template image to a given motion state and **adjoint** motion fields be warps from a given motion state back to the image template. Here we develop a multi-resolution reconstruction scheme that first solves for forward and adjoint low resolution motion fields and interpolates these motion fields to the desired resolution all as MSLR components. These interpolated motion fields at the desired resolution can then be further refined through k-space based template warping. These fields are then used in a final motion-compensated Extreme MRI reconstruction.

**Low Resolution Forward Motion Field Formulation**

Where applicable we follow the notation introduced in the MSLR reconstruction review above. Let acquired k-space data be stacked into a matrix $Y \in \mathbb{C}^{CM \times T}$. We model a bin of this time series in k-space as:

$$Y_t = \mathcal{A}(I_{temp}(\Omega_{for,t})) \quad (5)$$

where $I_{temp}$ is a template image, $\Omega_{for,t} \in \mathbb{R}^{3XN}$ represent 3 channel dense deformation fields of size $N$ with each voxel assigned a displacement: $\text{Id} + r(x, y, z)$ that warp the template image to a given motion state at time t. $\mathcal{A}$ is an operator that transforms this warped template image into k-space.

To both regularize the problem and fit data on the GPU, we represent the deformation fields $\Omega_{for,t}$ in a MSLR representation. Let $\boldsymbol{\Omega_{for}} \in \mathbb{R}^{3xTxN}$ be the spatiotemporal matrix of the stacked three channel deformation fields over $T$ frames. We decompose exactly as in (6) where:

$$\Omega_{for} = \sum_{j=1}^{J} \boldsymbol{B_j}(\Phi_{j,for}\Psi_{j,for}^H) \quad (6)$$

Where $\Phi_{j,for} \in \mathbb{R}^{3xNx1}, \Psi_{j,for}^H \in \mathbb{R}^{3xTx1}$ and $\boldsymbol{B_j}$ is the corresponding blocking operator.

Deformation fields are smoothed spatially using total variation regularization to allow for improved sliding motion at organ boundaries commonly found between the lung and chest wall (15). Although the MSLR representation significantly regularizes the deformation fields along the time dimension there is still potential for under-sampling artifact to propagate into the fields leading to high frequency oscillations through time in the image. To help mitigate this issue, block-wise rank of the MSLR deformation fields is minimized via the variational formulation of the nuclear norm. The regularization applied to the deformation field components at time $t$ is:

$$f_{reg,t} = \sum_{j=1}^{J} \frac{\lambda_j}{2} \left(\frac{1}{T}\|\Phi_{j,for}\|_F^2 + \|D\Psi_{j,for}^H\|_F^2\right) + \gamma\|D\Omega_{for,t}\| \quad (7)$$

Note that the regularization on $\Psi_{j,for}^H$ enforces temporal smoothness through the finite difference operator $D$ over time frames. The finite difference operator is also applied to compute approximate spatial gradients $(\frac{d\Omega}{dx}, \frac{d\Omega}{dy}, \frac{d\Omega}{dz})$ for total variation spatial smoothing of the deformation fields. The deformations fields are solved stochastically as in Ong et al (6). The complete objective function then to solve for forward motion fields at time $t$ in the MSLR basis is

$$\underset{\substack{\Phi_{j,for}, \Psi_{j,for}^H \\ \forall j \in J}}{\operatorname{argmin}} \|Y_t - \mathcal{A}(I_{temp}(\Omega_{for,t}))\| + \sum_{j=1}^{J} \frac{\lambda_j}{2}\left(\frac{1}{T}\|\Phi_{j,for}\|_F^2 + \|D\Psi_{j,for}^H\|_F^2\right) + \gamma\|D\Omega_{for,t}\| \quad (8)$$

**Low Resolution Adjoint Motion Field Formulation**

After solving for the forward motion fields, we solve for the adjoint motion fields that relate a motion state at time $t$ back to the template image. Forward motion fields are fixed and then applied to warp the chosen template frame to the motion state at time $t$. The MSLR representation of the adjoint deformation fields is then estimated by learning to warp this motion state back to the template. The algorithm then is:

*for iterations*

1. Randomly select time point $t \in \{t_1, t_2, \ldots, t_T\}$
2. Forward warp $I_{temp}$ to this motion state $I_{temp}(\Omega_{for,t})$
3. Optimize $\underset{\substack{\Phi_{j,adj}, \Psi^H_{j,adj} \\ \forall j \in J}}{\operatorname{argmin}} \|I_{temp} - I_{temp}(\Omega_{adj,t}(\Omega_{for,t}))\|^2 + \sum_{j=1}^{J} \frac{\lambda_j}{2}(\|\Phi_{j,adj}\|_F^2 + \|D\Psi^H_{j,adj}\|_F^2) + \gamma \|D\Omega_{adj,t}\|$

**MSLR Interpolation**

We then interpolate the MSLR representation of the low resolution forward and adjoint deformation fields to the desired resolution used for the final reconstruction. We first initialize $\Phi_{j,desired\ res}$ and $\Psi_{j,desired\ res}$ for the forward and adjoint fields that warp the time series at the desired resolution. The algorithm then is as follows

*for iterations:*

1. Randomly select time point $t \in \{t_1, t_2, \ldots, t_T\}$
2. Interpolate $\Omega_{low\ res,t} = \sum_{j=1}^{J} B_j(\Phi_{j,low\ res}\Psi^H_{j,low\ res})$ to $\Omega_{desired\ res,t} = \sum_{j=1}^{J} B_j(\Phi_{j,desired\ res}\Psi^H_{j,desired\ res})$ by applying a cubic B-spline interpolation operator
3. Optimize $\underset{\substack{\Phi_{j,desired\ res}, \Psi_{j,desired\ res} \\ \forall j \in J}}{\operatorname{argmin}} \|\Omega_{desired\ res,t} - \Omega_{low\ res,t}\|^2$

The interpolated motion fields at the desired resolution can then be further refined by the same k-space based motion field estimation introduced earlier.

**Motion Compensated Extreme MRI**

We then integrate the MSLR representation of the forward and adjoint motion fields that warp the time series at the desired resolution into Extreme MRI.

$$\min_{L_j, R_j\ \forall j \in J} \|Y_t - \mathcal{A}[I_t(\Omega_{for,t})]\|^2 + \sum_{j=1}^{J} \frac{\lambda_i}{2}(\frac{1}{T}\|L_j\|_F^2 + \|R_j\|_F^2) \quad (9)$$

Where $I_t = \sum_{j=1}^{J} M_j(L_j R^H_{j,t})$ and $\Omega_{for,t} = \sum_{j=1}^{J} M_j(\Phi_{j,t}\Psi^H_{j,t})$

The algorithm using stochastic gradient descent proceeds as follows:

Initialize $\{L_j\}_{j=1}^{J}$ and $\{R_j\}_{j=1}^{J}$ as in (6) then

*for iterations:*

1. Randomly choose a time frame $t$ and reconstruct its image: $I_t = \sum_{j=1}^{J} M_j(L_j R_{j,t})$, and associated forward and adjoint fields: $\Omega_{for,t} = \sum_{j=1}^{J} M_j(\Phi_{for,j,t} \Psi_{for,j,t}^H)$, $\Omega_{adj,t} = \sum_{j=1}^{J} M_j(\Phi_{adj,j,t} \Psi_{adj,j,t}^H)$. $I_t$ should be aligned with all other time frames.
2. Warp this image to its appropriate motion state: $I(\Omega_{for,t})$
3. Take the gradients of the data-consistency term with respect to $\{L_j\}$ and $\{R_{j,t}\}$. By the chain rule first take the gradient of the data-consistency term: $DC_{grad}$ with respect to $I_t(\Omega_{for,t})$, warp this gradient back to the aligned space using the adjoint deformation field: $DC_{grad}(\Omega_{adj,t})$, and finally take the gradient with respect to $\{L_j\}$ and $\{R_{j,t}\}$.
4. Take the gradients of $f_{reg} = \sum_{j=1}^{J} \frac{\lambda_i}{2} (\|L_j\|_F^2 + \|R_{j,t}\|_F^2)$ with respect to $\{L_j\}$ and $\{R_{j,t}\}$.
5. Update L and R as follows: $L_j = L_j - \alpha T[\nabla_{L_j} f_{reg} - \nabla_{L_j}(DC_{grad}(\Omega_{adj}))]$ and $R_{j,t} = R_{j,t} - \alpha[\nabla_{R_j} f_{reg} - \nabla_{R_j}(DC_{grad}(\Omega_{adj}))]$

## 3. Methods

We applied MoCo-MSLR to free breathing 3D radial imaging acquisitions in the lung and placenta from previously acquired datasets. Lung data was acquired in one healthy volunteer and 2 patients with diffuse lung disease [cystic fibrosis (CF) and idiopathic pulmonary fibrosis (IPF)]. Placental data was acquired in one healthy pregnant patient in the third trimester. All subjects were asked to breath normally during the acquisition. For all subjects, reconstructions at ~500ms were performed to resolve respiratory motion. For subjects with sufficient contrast between the ventricular wall and blood (healthy volunteer and CF case), a second reconstruction at ~100ms was run to resolve both cardiac and respiratory motion.

### 3.1 Reconstruction Implementation

K-space data was coil compressed to 20 channels if greater than 20 channels were used during acquisition, otherwise data was not coil compressed. Similar to (6), the 3D radial data used an oversampled field of view (FOV) and was adjusted automatically to include all areas producing MRI signal. Signal outside the reconstructed FOV can lead to artifacts from data-inconsistencies between the acquired k-space data and the NUFFT transformed image data. Further, modeling motion that falls in and out of the FOV is difficult and leads to non-topology preserving deformation fields. To counter this, we followed the steps in (6) by reconstructing a gridded image at twice the prescribed FOV, thresholding the image at 0.1 of the maximum amplitude to estimate the FOV. Density compensation was used to improve convergence. Sensitivity maps were estimated using J-sense from all data binned together (16). For the motion correction steps that require k-space data and the final MSLR reconstruction, k-space data was binned in time with number of projections per bin determined by dividing the total number of projections by the number of required frames for reconstruction.

Low resolution template images (~3.5 mm isotropic) were reconstructed by running an Extreme MRI reconstruction with all projections binned together. The reconstruction was run for 200 iterations to ensure data-consistency. Block sizes of [8,16,32] with regularization weight of 1e-8 were used across all cases, however, these choices do not substantially impact the template reconstruction as only a single frame was reconstructed.

Spatial deformation field bases $\{\Phi_j\}_{j=1}^{J}$ were initialized using Gaussian noise and temporal deformation field bases $\{\Psi_j^H\}_{j=1}^{J}$ were initialized with all 0s.

In place of explicitly computing gradients for the low-resolution motion estimation and interpolation steps, we used auto differentiation in Pytorch using an Adam optimizer. For low resolution steps, a learning rate of .01 across all block scale was chosen. For interpolation, a learning rate of .001 across all block scales was chosen. To fit the spatial deformation field bases used in the full resolution reconstruction with matrix size $P_x \, x \, P_y \, x \, P_z$ on the GPU, we created blocks corresponding to a matrix of control points of size $\frac{Px}{3} \, x \, \frac{Py}{3} \, x \, \frac{Pz}{3}$ that was then trilinearly interpolated to the full deformation field size during reconstruction.

The final motion compensated reconstruction used the code found at https://github.com/mikgroup/extreme_mri as a foundation. This code was modified to allow for forward and adjoint warping of time frames. For all MoCo-MSLR reconstructions, we represented the time series using 2 block scales with sizes [64,128] to allow the reconstructions to fit on the GPU.

For all MoCO-MSLR reconstructions, image quality and motion dynamics were compared against Extreme MRI. For all Extreme MRI reconstruction, three block scales with block-sizes of [32,64,128] with regularization weight of 1e-8 were used. These reconstructions were run for 60 iterations. For reconstructions with targeted temporal resolution ~500ms, respiratory dynamics was tracked by fixing a volumetric window about the liver-lung interface, and then auto-correlating this fixed window with a sliding window through time. For reconstructions with targeted temporal resolution near ~100ms, both cardiac and respiratory dynamics were tracked if the motion was resolved on visual inspection of CINEs. Cardiac dynamics was tracked by fixing a volumetric window about the left ventricle, autocorrelating as above, Fourier transforming this signal, and then filtering the signal in a 0.05hz pass band about the presumed cardiac cycle rate.

Respiratory dynamics was tracked as above and then gaussian smoothed using $\sigma = 3$ pixels in Scipy.

### 3.2 Healthy Volunteer 1

One healthy volunteer UTE lung dataset (17) was acquired with a 32 channel coil, scan time of 5 minutes and 45 seconds, TE=0.25ms, TR=3.6ms, flip angle=24° and 1.25mm isotropic resolution, Ferumoxytol (4mg/kg) was given prior to the scan. The number of projections was 94,957 with 636 readout length acquired using 3D pseudorandom bit-reversed view ordering. Two reconstructions were performed. The first reconstruction targeted a spatial and temporal resolution of 1.25mm isotropic and 690ms with the goal of resolving respiratory motion. The second reconstruction targeted a spatial and temporal resolution of 1.67mm isotropic and 115ms respectively with the goal of resolving both cardiac and respiratory motion.

### 3.3 Cystic Fibrosis Patient

One UTE lung dataset of a cystic fibrosis (CF) patient was acquired with an 8-channel coil array, an overall scan time of 4 minutes 18 seconds, TE=80µs, TR=3.48ms, flip angle 4 degrees and 1.25 mm isotropic resolution. The number of projections was 75,768 and 654 readout length. This dataset is publicly available and was included in the original Extreme MRI work [1]. Two reconstructions were performed. The first reconstruction targeted a spatial and temporal resolution of 1.25mm isotropic and 515ms respectively with the goal of resolving respiratory motion. The second reconstruction targeted a spatial and temporal resolution of 1.67 mm isotropic and 83ms temporal respectively with the goal of resolving both cardiac and respiratory motion.

### 3.4 IPF Patient

One UTE lung dataset of a patient with idiopathic pulmonary fibrosis (IPF) was acquired with an 8-channel coil array, an overall scan time of 4 minutes 54 seconds, TE=80µs, TR=3.27ms, flip angle 4 degrees and 1.25 mm isotropic resolution. The number of projections was 89964 and 654 samples per projection. One reconstruction was performed. The targeted spatial and temporal resolution for this reconstruction was 1.25mm isotropic and 588ms respectively with the goal of resolving respiratory motion.

### 3.5 Third Trimester Pregnant Patient

One placental dataset of a healthy pregnant patient in the third trimester was acquired with GE Air Coil, an overall scan time of 4 minutes, 2 seconds, TE=1.3ms, TR=5.0ms, flip angle of 25 degrees, 1mm isotropic resolution. One reconstruction was performed. The targeted spatial and temporal resolution for this reconstruction was 1.8 mm isotropic and 605ms with the goal of resolving respiratory motion.

The healthy volunteer and CF datasets were acquired on a 3 Tesla GE scanner. The IPF and placental datasets acquired on a 1.5 Tesla GE scanner

## 4. Results

**Figure 1** shows extracted respiratory signals for ~500 ms reconstructions across all cases.

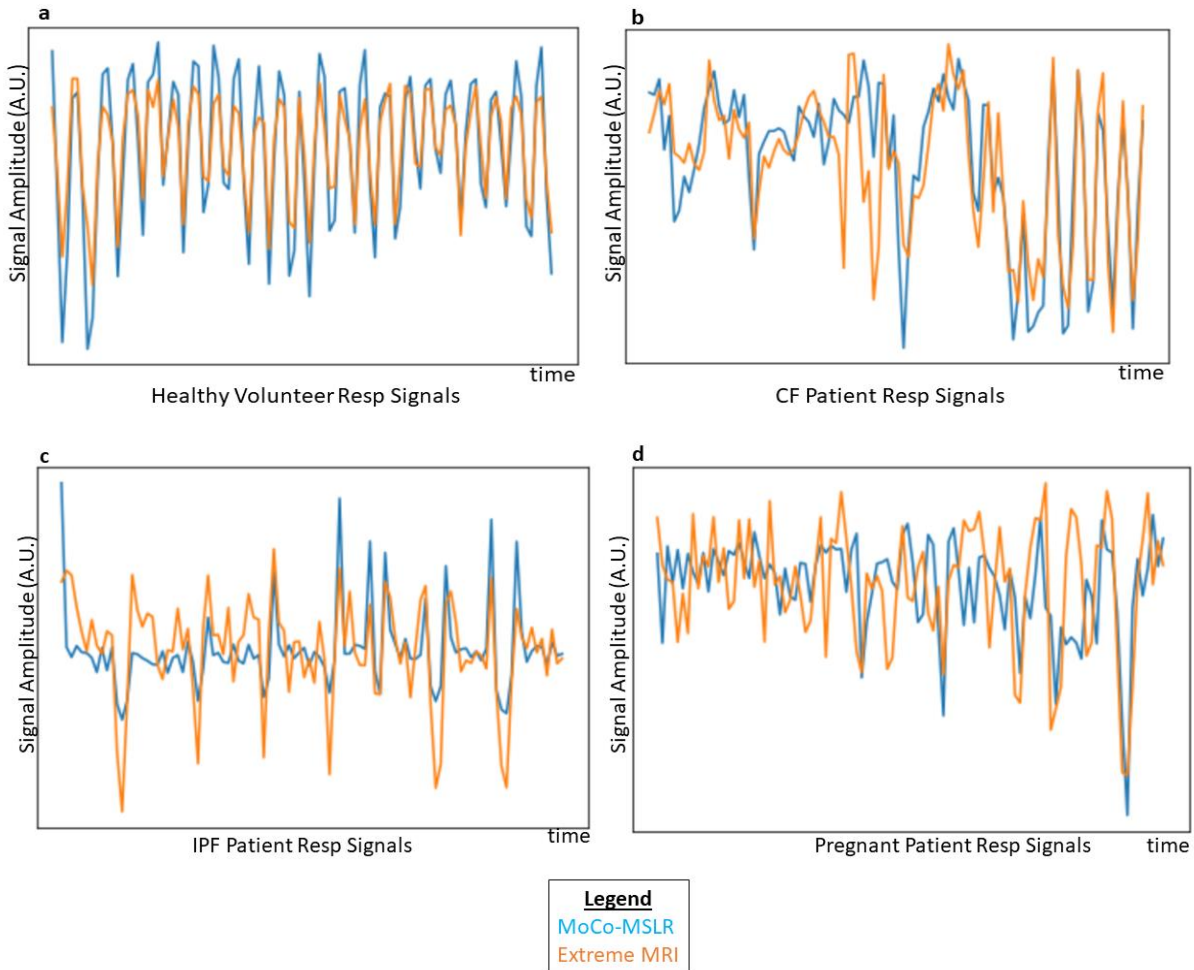

*Figure 1: Respiratory Signal Tracking. For reconstructions near 500ms that visualized the diaphragm, a fixed volumetric window was placed on the right hemidiaphragm and autocorrelated with a sliding window through time. The respiratory dynamics in the (**a**) healthy volunteer is nearly periodic. Both MoCo-MSLR and Extreme-MRI are in phase. The respiratory dynamics in the (**b**) CF patient were much more variable. In general though, MoCo-MSLR and Extreme-MRI are roughly in phase. In the (**c**) IPF patient, respiratory dynamics between MoCo-MSLR and Extreme-MRI are generally in phase. In the (**d**) pregnant patient, a fixed volumetric window was placed on the edge between the uterine wall and placenta and autocorrelated with a sliding window through time. Although respiratory dynamics are a little harder to extract here, overall, both MoCo-MSLR and Extreme-MRI remain roughly in phase.*

## 4.1 Healthy Volunteer Dataset

**Figure 2** and **supplemental video 1** (https://doi.org/10.6084/m9.figshare.19583887.v2) compare MoCo-MSLR versus Extreme MRI for the reconstruction targeting 690ms temporal resolution.

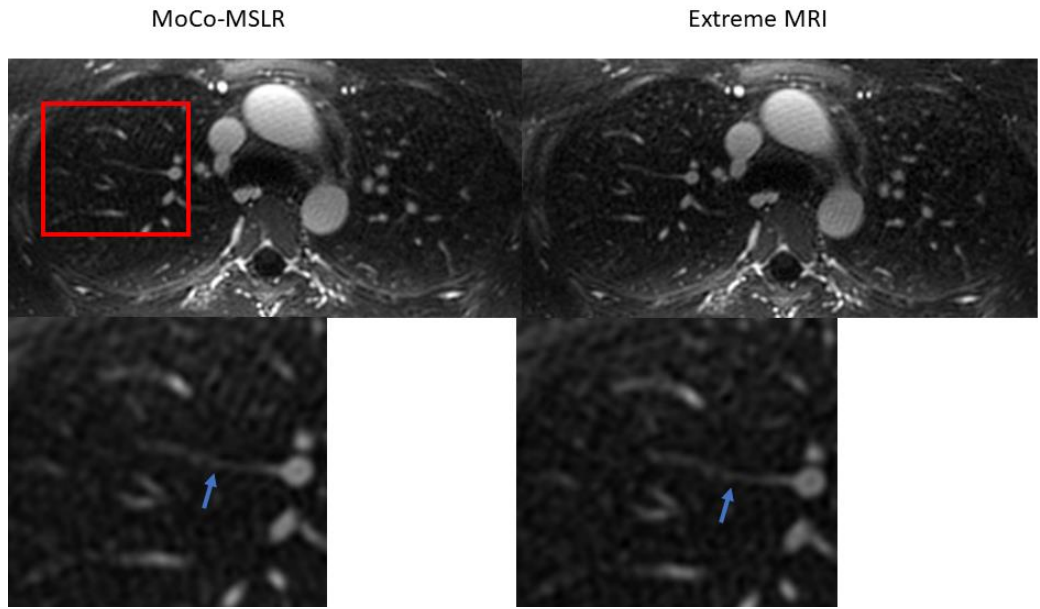

*Figure 2: Reconstruction Results on Healthy Volunteer. Displayed here are representative axial slices from MoCo-MSLR (left) and Extreme MRI (right) reconstructions with 690ms targeted temporal resolution and 1.25mm isotropic spatial resolution. The red bounding box represents the portion of the image zoomed in on row 2. In this healthy volunteer with nearly periodic respiratory motion, no significant differences in image quality can be seen. Both reconstructions resolve small vascular features equally well (blue arrow)*

Image quality is similar between the reconstruction methods with minimal flickering artifact; however, the liver edge appears sharper for MoCo-MSLR during motion **(supplemental video 1)**. Vascular structures are resolved similarly by both methods **(figure 2, blue arrow, row 2).** Both reconstructions resolve similar motion dynamics as seen from the video and the extracted respiratory signal **(figure 1a).**

**Figure 3** and **supplemental video 2 (https://doi.org/10.6084/m9.figshare.19583914.v1** ) compare MoCo-MSLR versus Extreme MRI for the reconstruction targeting 115ms temporal resolution. From **supplemental video 2**, MoCo-MSLR resolves cardiac and respiratory dynamics. Respiratory dynamics and some degree of left ventricular wall motion are resolved by Extreme MRI. Significant blurring though at both the diaphragm and left lateral ventricular wall is observed. MoCo-MSLR shows limited blurring of these structures.  Similar findings can be seen in **figure 3a and 3b**.

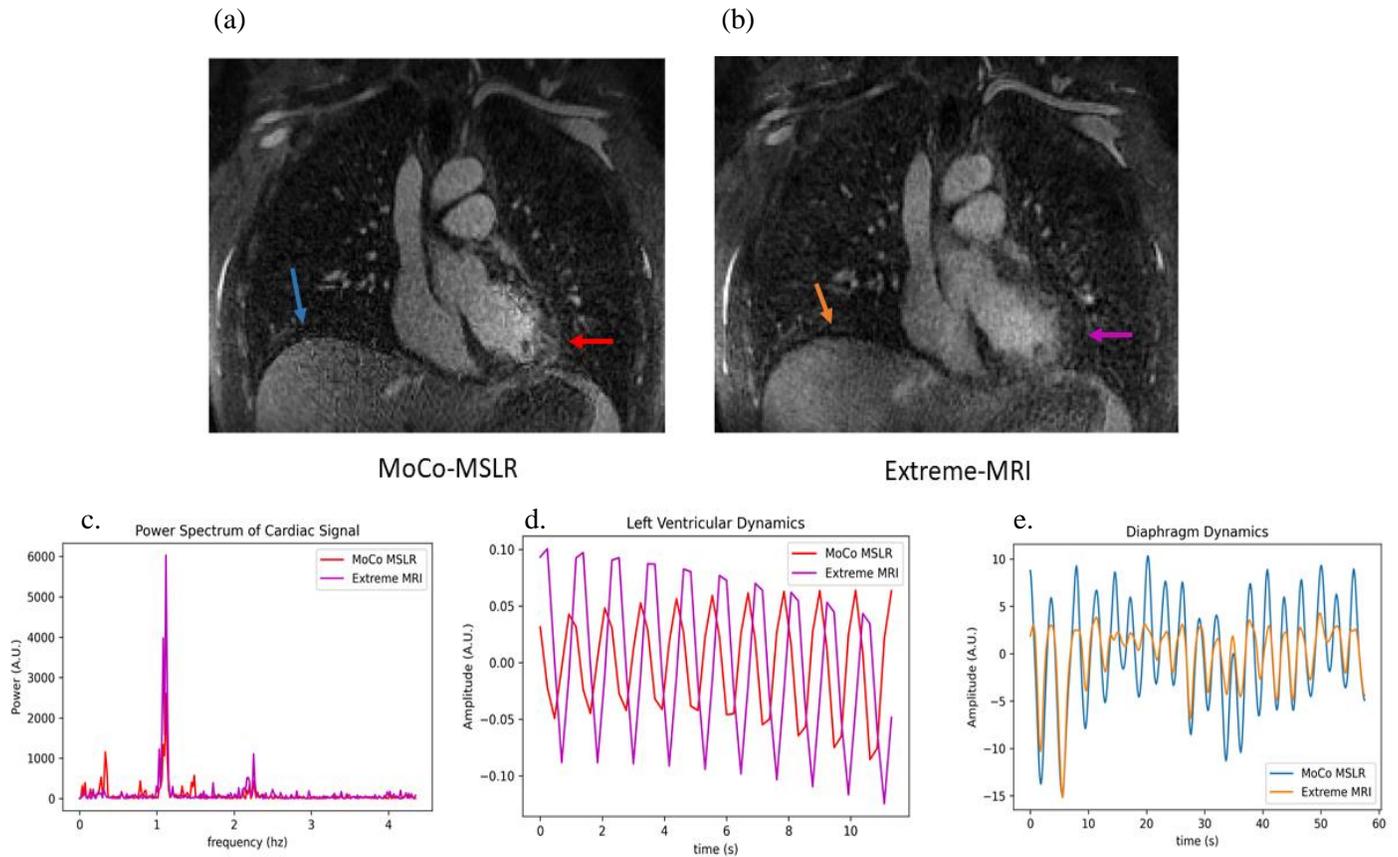

*Figure 3:* *Cardiac and Respiratory Dynamics at High temporal resolution. MoCo-MSLR and Extreme MRI reconstructions were run on the healthy volunteer at a targeted temporal resolution of 115ms and spatial resolution of 1.67 mm isotropic. Two volumetric windows were fixed about the lateral left ventricular wall (red/purple arrows) and the right hemidiaphragm (blue/orange arrows), and autocorrelated with a sliding window at the same spatial location through time to extract cardiac and respiratory dynamics respectively. The power spectrum (**c**) of the autocorrelation about the lateral left ventricular wall was then computed demonstrating a strong frequency peak around 1.11 hz corresponding to a physiologically reasonable 68 beats per minute. Cardiac signal (**d**) was then extracted by filtering a 0.05hz passband around the peak signal in frequency space. Both MoCo-MSLR and Extreme-MRI cardiac signals maintain the same phase relationship through time. The autocorrelation around the right hemidiaphragm was Gaussian smoothed to reveal respiratory dynamics (**e**). Both reconstructions remain in the same respiratory phase through multiple respiratory cycles. It is important to note from **supplemental video 2** that the cardiac motion resolved in MoCo-MSLR is more realistic than that resolved by Extreme MRI. Evidence for this can be seen comparing the sharpness of (**a**) and (**b**) about the diaphragm (blue/orange arrow) and lateral left ventricle (red/purple arrow). In both locations, MoCo-MSLR is significantly sharper than Extreme-MRI.*

Although the cardiac dynamics in **supplemental video 2** in the MoCo-MSLR reconstruction appear much more realistic than in Extreme MRI, both methods demonstrate strong peaks in their Fourier power spectra at 1.11hz corresponding to a heart rate of 68 beats/min (**figure 3c**). Filtering this signal in a small passband around this frequency results in signals that resemble cardiac waveforms (**figure 3d**). Diaphragm dynamics **(figure 3e)** also appear to be in phase.

**Supplemental video 3** (https://doi.org/10.6084/m9.figshare.19583932.v3) demonstrates axial, 2 chamber, 4 chamber, and short axis views of heart for the MoCo-MSLR reconstruction. Multiple cardiac phases in all views are clearly captured. **Figure 4, row 1** demonstrates left ventricular phases from late diastole to systole for the healthy volunteer (MRA)

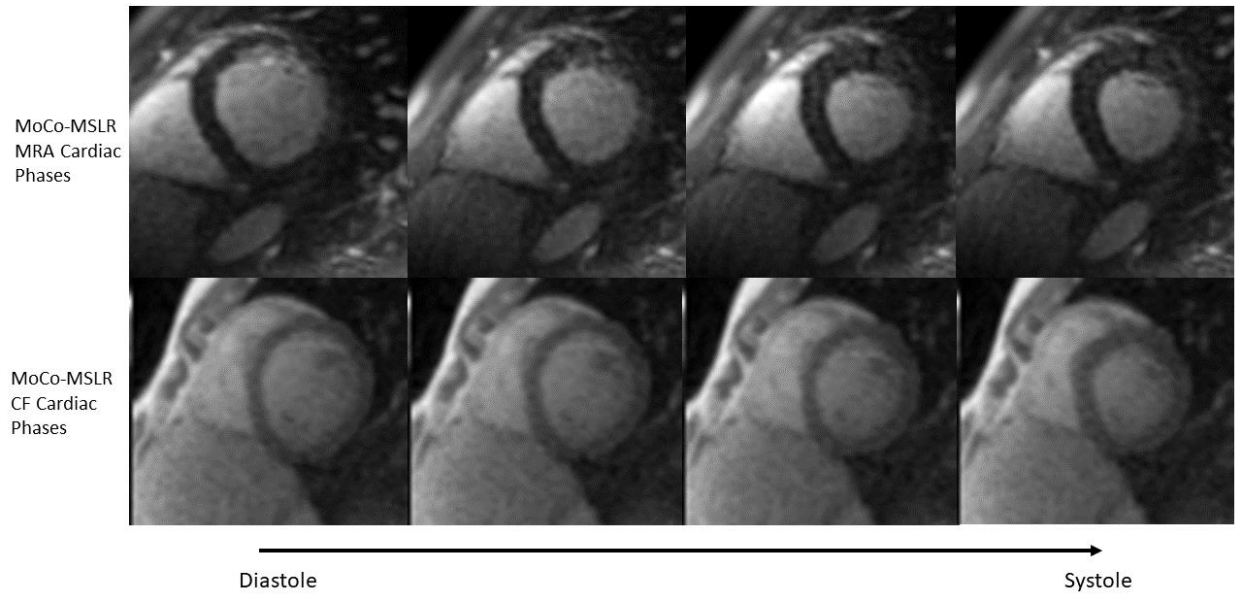

*Figure 4: Short axis Cardiac Phases. Cardiac dynamics from mid/late diastole through systole are shown from MoCo-MSLR on the healthy volunteer (targeted temporal resolution: 115ms) and patient with cystic fibrosis (targeted temporal resolution: 83ms)*

## 4.2 Cystic Fibrosis Lung Dataset

**Figure 5** and **supplemental video 4** (https://doi.org/10.6084/m9.figshare.19583938.v1) compares MoCo-MSLR versus Extreme MRI for the reconstruction targeting 515ms temporal resolution.

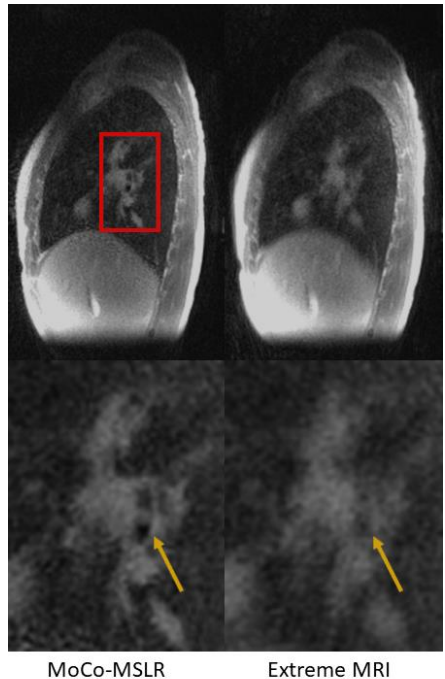

*Figure 5: Reconstructions results on Patient with Cystic Fibrosis. Displayed here are representative sagittal slices from both reconstructions (targeted temporal resolution: 515 ms, spatial resolution: 1.25 mm isotropic). In the zoomed-out images in row 1, MoCo-MSLR sharply resolves the liver edge and larger airway structures compared to Extreme MRI. This can be seen even more clearly in the zoom-in images on row 2 (orange arrow).*

**Figure 5** shows that the MoCo-MSLR is significantly sharper than Extreme MRI demonstrating airway feature blurred out in Extreme MRI (yellow arrow). Similar findings are seen in **supplemental video 4** where significant blurring of the liver edge and small vascular structures are seen in the Extreme MRI reconstruction. These structures remain sharp for MoCo-MSLR. From the extracted respiratory signal alone **(figure 1b),** motion dynamics are similar. However, bulk motion and tracheal collapse seen in the MoCo-MSLR reconstruction are not observed in the Extreme-MRI reconstruction (**supplemental video 4**).

**Figure 6** and **supplemental video 5 (https://doi.org/10.6084/m9.figshare.19583944.v2)** compare MoCo-MSLR versus Extreme MRI for the reconstruction targeting 83ms temporal resolution. MoCo-MSLR does resolve cardiac and respiratory dynamics, however, high frequency oscillations through time are present. Further, significant flickering artifact is observed. No obvious left ventricular wall motion is seen in the Extreme MRI reconstruction. Some small motions at the diaphragm are seen, however this is partly obscured by blur.

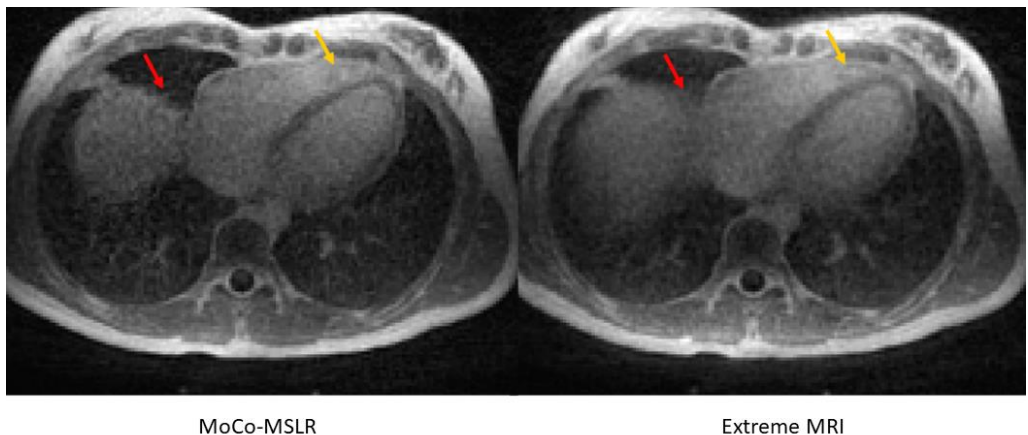

*Figure 6*: *Reconstruction Results on Patient with Cystic Fibrosis at high temporal resolution. Displayed here are representative axial slices from both reconstructions (targeted temporal resolution: 83ms, spatial resolution: 1.67 mm isotropic). Moco-MSLR is sharper particularly around structures like the liver that should be in motion due to respiration (red arrow) and the heart. Note that some subtle non-physiologic warping over the heart (yellow) can be seen in the MoCo-MSLR reconstruction.*

**Figure 6** shows that the MoCo-MSLR reconstruction has reduced blur around the heart relative to Extreme MRI (red arrow). However, some non-physiologic warping can be seen in the MoCo-MSLR reconstruction near the anterior part of the cardiac septum (yellow arrow). Comparisons between the dynamics for these two reconstructions were not performed as no cardiac dynamics and only subtle diaphragm motion was seen in Extreme MRI. **Supplemental video 6 (https://doi.org/10.6084/m9.figshare.19583950.v2)** is a 15 frame CINE of axial, 2 chamber, 4 chamber, and short axis views of the heart again demonstrating realistic cardiac dynamics in all views. High frequency oscillations can clearly be seen. **Figure 4** (**row 2**) demonstrates left ventricular phases from late diastole to systole for the CF patient.

## 4.3 Idiopathic Pulmonary Fibrosis Dataset

**Figure 7** and **supplemental video 7(https://doi.org/10.6084/m9.figshare.19583953.v1)** compare MoCo-MSLR versus Extreme MRI for IPF reconstructions targeting 588ms temporal resolution. From **supplemental video 7**, structures around the lung hilum are sharp for MoCo-MSLR throughout respiration. These structures are blurred somewhat in Extreme MRI. Additionally, there is less flickering artifact in the MoCO-MSLR reconstruction than Extreme MRI. Notice that the blur around the liver edge in Extreme MRI is replaced by warping artifact in MoCo-MSLR.

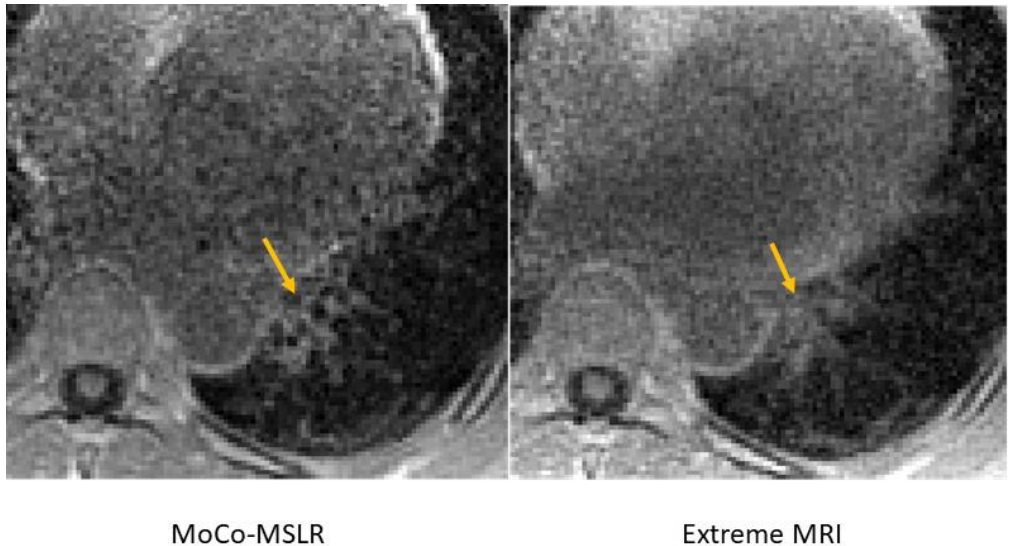

*Figure 7:* *Reconstructions Results on IPF Patient. Representative axial slices near the lung base are shown for both reconstructions. Fibrosis around the airway and the airways themselves are more clearly resolved in MoCo-MSLR than Extreme MRI.*

In **figure 7**, MoCo-MSLR clearly resolves small airways and associated fibrosis (orange arrow) not visualized in Extreme MRI. From both **supplemental video 7** and **figure 1c**, it appears that respiratory motion is similar between the two reconstructions, however, the MoCo-MSLR reconstruction does appear to miss a transient diaphragm excursion seen in Extreme MRI. **Supplemental video 8 (https://doi.org/10.6084/m9.figshare.19583956.v1)** shows a sagittal slice paired with its associated motion field through time demonstrating how the displacement field changes throughout the respiratory cycle.

### 4.4 Third Trimester Pregnant Patient Dataset

**Figure 8** and **supplemental video 9** (https://doi.org/10.6084/m9.figshare.19583959.v1) compare MoCo-MSLR and Extreme MRI for reconstructions targeting 605ms temporal resolution

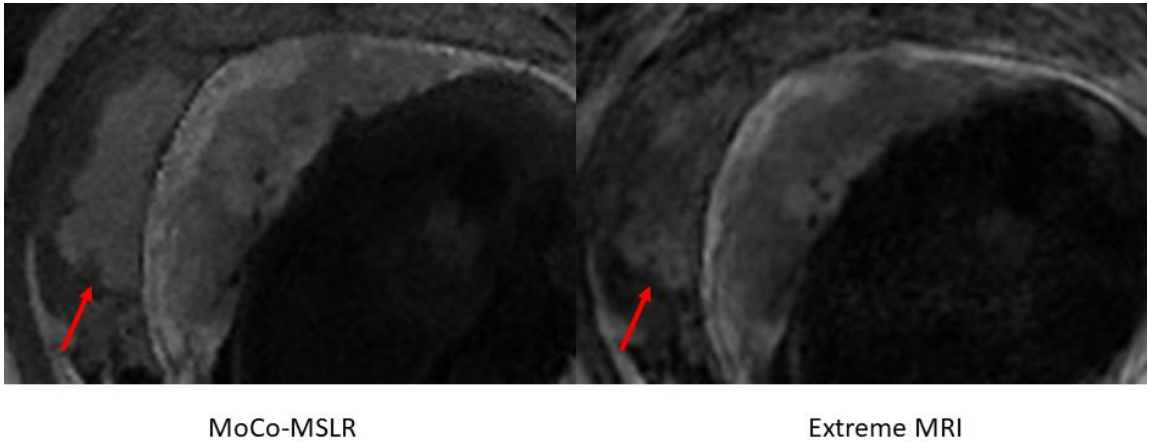

*Figure 8: Reconstructions Results on Healthy Pregnant Patient in Third Trimester. View showing the placenta and uterine layers. Significant artifact obstructs the uterus not seen in MoCo-MSLR (red arrow).*

**Figure 8** shows that MoCo-MSLR results in sharper delineation between uterine layers than Extreme MRI where these layers are obscured by artifact. Motion dynamics appear to be similar between reconstructions both in **supplemental video 9** and from the respiratory signals in **figure 1d**. A uterine contraction is observed from 3.38 to 5.93 time units.

### 5. Discussion

In this work, we developed a method called MoCo-MSLR to estimate and then integrate memory efficient representations of forward and adjoint motion deformation fields into Extreme MRI reconstructions. In MoCo-MSLR, low resolution motion fields are first learned directly as multiscale low rank components by enforcing k-space based loss between a deformed template and acquired k-space data. These fields are then interpolated in the MSLR space to match the desired full resolution reconstruction. Finally, the deformation fields and their adjoint are incorporated into Extreme MRI in the forward model. By using compact representations for both motion fields and the time series, motion compensated high spatiotemporal reconstructions are made possible with very low memory footprint.

MoCo-MSLR results in improved image quality compared to Extreme MRI at ~500ms temporal resolution. Image quality improvements seen with our method include reduced undersampling and flickering artifacts, sharper image features, the ability to resolve small vascular and airway features, and resolve certain dynamics not seen in Extreme MRI reconstruction. MoCo-MSLR at higher temporal resolutions (~100ms) realistically captures cardiac dynamics. Extreme MRI incompletely resolved cardiac dynamics in the healthy volunteer with high blood pool to myocardium contrast. In the CF case with lower blood pool to myocardium contrast, Extreme MRI completely failed to resolve cardiac dynamics.

Our work demonstrates similar image quality improvement seen with past strategies incorporating motion fields directly into reconstructions. This improvement was expected to some extent because the time series modeled by the left spatial and right temporal bases in MoCo-MSLR is aligned meaning maximal correlations exist across frames. Image quality improvements can be seen in the work of (5,12) when aligning data during reconstruction. Our model simply extends this notion of improved reconstruction through alignment to a much larger scale problem. Without motion correction, the left spatial and right temporal bases in MSLR model all dynamics in the time series which reduces the degree of correlation across frames ultimately reducing image quality.

There were, however, significant variations in the degree of image quality improvement across cases. This appeared to be, in part, related to the complexity of motion. These differences can be seen particularly well when comparing the healthy volunteer with nearly periodic motion (**supplemental video 1**) to the CF patient with both irregular respiratory and bulk motions. In the healthy volunteer, MoCo-MSLR and Extreme MRI are comparable with respect to image quality (**supplemental video 1** and **figure 2**) at temporal resolution targeting respiratory motion (~500ms). Minimal flickering and streaking artifact are seen, and small vascular features are resolved well by both reconstruction methods. On the other hand, MoCo-MSLR demonstrated significantly higher image quality (**supplemental video 4** and **figure 5**) then the Extreme MRI reconstruction for the CF patient. The liver edge is sharper in MoCo-MSLR even during irregular respiratory motion. Additionally, airway/vascular features blurred out in Extreme MRI are clearly resolved in the MoCo-MSLR reconstruction (**figure 5**). One possible explanation for this is Extreme MRI is not actually resolving all motion at the targeted temporal resolution which would lead to blur. For instance, in **supplemental video 4**, bulk motions and tracheal collapse seen in the MoCo-MSLR reconstruction are not observed in Extreme MRI. Although there is no way to validate if these motions are real, the quality of the MoCo-MSLR reconstructions suggests they are. Further, tracheomalacia which can lead to tracheal collapse especially when there are large fluctuations in thoracic pressures (e.g., during a cough) is common in patients with cystic fibrosis.

In general, MoCo-MSLR does appear to resolve irregular respirations and bulk motion with minimal blurring better than Extreme MRI. This makes sense because as mentioned in (6), irregular respirations and bulk motion are not necessarily low rank even for small block sizes. By explicitly modeling these motions, MoCo-MSLR significantly reduces blur while capturing these motions. A counterargument to this is motion fields represented using multi-scale low rank components may suffer from the same issue. Although to some extent this is true, deformation fields only have to model motion, not the background plus dynamics and thus may admit more compressible representations allowing MoCo-MSLR to reconstruct even more undersampled data with high fidelity than the original Extreme MRI approach. The ability of MoCo-MSLR to capture cardiac dynamics at ~100ms temporal resolution while Extreme MRI struggles lends experimental evidence to this hypothesis.

At high temporal resolutions (~100ms) significant differences in reconstruction quality remain both when comparing MoCo-MSLR to Extreme MRI and when comparing each reconstruction to itself across different cases. Although complexity of motion may still play a role here, it appears that higher SNR results in improved ability to capture high temporal resolution dynamics. This can be seen when comparing the higher SNR contrast enhanced healthy volunteer acquisition to the lower SNR CF acquisition. Extreme MRI captures some cardiac motion in the healthy volunteer, but no cardiac motion can be seen in the lower SNR CF acquisition. Although MoCo-MSLR captures cardiac dynamics in both the healthy volunteer and CF patient, the CF reconstruction has significantly more high frequency oscillations present (**supplemental video 5**) suggesting the deformation fields are also modeling noise in addition to signal. This preliminary finding suggests that at high temporal resolution, contrast-enhanced acquisitions may be preferred.

There are a number of limitations to this work. There are several image artifacts that arise because the deformation fields are not topology preserving (i.e. non-diffeomorphic). In the IPF case (**supplemental video 7**), a sandpaper like texture can be seen in and around the liver edge. In L.T's experience using other motion correction algorithms like iMoCo, these same artifacts arise when the deformation fields are not topology preserving i.e. non-diffeomorphic. Use of algorithms that ensure the fields are diffeomorphic removes these artifacts in the context of iMoCo. A related warping artifact can be seen in the high temporal resolution reconstructions. This artifact occurs when tissues that locally should be moving together, displace with different velocities essentially tearing the tissue apart. The result is a kind of blurring. A potential direction for this work is to develop multi-scale compressed representations for diffeomorphic fields. It is not immediately clear though how to develop such a method with theoretical guarantees.

Another artifact unrelated to non-diffeomorphic fields seen primarily in the ~100ms resolution is high frequency oscillations. This is significantly worse in the CF case then the healthy volunteer with the same regularization weights. Although the regularization on both spatial smoothing and rank minimization can be increased to attempt to remove this artifact, the higher the regularization weight, the more difficult it becomes to capture motion. Exploring the hypothesis that ability to resolve high temporal resolution dynamics may be dependent on SNR may be fruitful to better define acquisition parameters to generate optimal high temporal resolution reconstructions.

Similar to (6), it is unknown whether the prescribed temporal resolution matched the true dynamics at that temporal resolution. Validation is a major challenge for this work. Few real time imaging modalities can scan simultaneously with MR to provide ground truth data, however, recent progress in simultaneous MRI/Ultrasound systems (18) may be a promising future approach for validation.

Finally, in its current form, MoCo-MSLR only works for images without contrast dynamics as it relies on warping a fixed template. The ability to incorporate motion estimation for high spatiotemporal reconstruction of acquisitions with contrast dynamics is an interesting avenue for future work.

## 6. Conclusion:

In this work we improve on a state-of-the-art image reconstruction algorithm (Extreme MRI) by incorporating motion fields into the reconstruction. We demonstrate that MoCo-MSLR makes it possible to reconstruct motion compensated 3D dynamic acquisitions at high spatiotemporal resolutions in a computationally efficient manner. Our method shows improved image sharpness and motion robustness when compared to Extreme MRI at the same temporal resolution. Additionally, when pushed to temporal resolutions of ~100ms, MoCo-MSLR can depict cardiac and respiratory dynamics beyond the capabilities of Extreme MRI.

## 7. Acknowledgements

The authors wish to acknowledge support from NIH R01CA190298, NIH/NHLBI HL 136965, and GE Healthcare.

## 8. References

1. Jaimes C, Kirsch JE, Gee MS. Fast, free-breathing and motion-minimized techniques for pediatric body magnetic resonance imaging. Pediatr Radiol. 2018 Aug 1;48(9):1197–208.

2. Zhu X, Tan F, Johnson K, Larson P. Optimizing trajectory ordering for fast radial ultra-short TE (UTE) acquisitions. J Magn Reson. 2021 Jun 1;327:106977.

3. Zhang T, Cheng JY, Potnick AG, Barth RA, Alley MT, Uecker M, et al. Fast Pediatric 3D Free-breathing Abdominal Dynamic Contrast Enhanced MRI with High Spatiotemporal Resolution. J Magn Reson Imaging JMRI. 2015 Feb;41(2):460–73.

4. Torres L, Kammerman J, Hahn AD, Zha W, Nagle SK, Johnson K, et al. "Structure-Function Imaging of Lung Disease Using Ultra-Short Echo Time MRI." Acad Radiol. 2019 Mar;26(3):431–41.

5. Zhu X, Chan M, Lustig M, Johnson KM, Larson PEZ. Iterative motion-compensation reconstruction ultra-short TE (iMoCo UTE) for high-resolution free-breathing pulmonary MRI. Magn Reson Med. 2020;83(4):1208–21.

6. Ong F, Zhu X, Cheng JY, Johnson KM, Larson PEZ, Vasanawala SS, et al. Extreme MRI: Large-scale volumetric dynamic imaging from continuous non-gated acquisitions. Magn Reson Med. 2020;84(4):1763–80.

7. Ludwig J, Speier P, Seifert F, Schaeffter T, Kolbitsch C. Pilot tone–based motion correction for prospective respiratory compensated cardiac cine MRI. Magn Reson Med. 2021;85(5):2403–16.

8. Feng L, Axel L, Chandarana H, Block KT, Sodickson DK, Otazo R. XD-GRASP: Golden-Angle Radial MRI with Reconstruction of Extra Motion-State Dimensions Using Compressed Sensing. Magn Reson Med. 2016 Feb;75(2):775–88.

9. Otazo R, Candès E, Sodickson DK. Low-rank plus sparse matrix decomposition for accelerated dynamic MRI with separation of background and dynamic components. Magn Reson Med. 2015;73(3):1125–36.

10. Guaranteed Minimum-Rank Solutions of Linear Matrix Equations via Nuclear Norm Minimization [Internet]. [cited 2022 Apr 12]. Available from: https://epubs.siam.org/doi/epdf/10.1137/070697835

11. Burer S, Monteiro RDC. A nonlinear programming algorithm for solving semidefinite programs via low-rank factorization. Math Program. 2003 Feb 1;95(2):329–57.

12. Otazo R, Koesters T, Candes E, Sodickson DK. Motion-guided low-rank plus sparse (L+ S) reconstruction for free-breathing dynamic MRI. In: Proc Intl Soc Mag Reson Med. 2014. p. 0742.

13. Huttinga NRF, Bruijnen T, van den Berg CAT, Sbrizzi A. Nonrigid 3D motion estimation at high temporal resolution from prospectively undersampled k-space data using low-rank MR-MOTUS. Magn Reson Med. 2021;85(4):2309–26.

14. Ong F, Lustig M. Beyond low rank + sparse: Multi-scale low rank matrix decomposition. In: 2016 IEEE International Conference on Acoustics, Speech and Signal Processing (ICASSP). 2016. p. 4663–7.

15. Vishnevskiy V, Gass T, Székely G, Goksel O. Total Variation Regularization of Displacements in Parametric Image Registration. In: Yoshida H, Näppi JJ, Saini S, editors. Abdominal Imaging Computational and Clinical Applications [Internet]. Cham: Springer International Publishing; 2014 [cited 2022 Apr 12]. p. 211–20. (Lecture Notes in Computer Science; vol. 8676). Available from: http://link.springer.com/10.1007/978-3-319-13692-9_20

16. Ying L, Sheng J. Joint image reconstruction and sensitivity estimation in SENSE (JSENSE). Magn Reson Med. 2007;57(6):1196–202.

17. Knobloch G, Colgan T, Schiebler ML, Johnson KM, Li G, Schubert T, et al. Comparison of gadolinium-enhanced and ferumoxytol-enhanced conventional and UTE-MRA for the depiction of the pulmonary vasculature. Magn Reson Med. 2019 Nov;82(5):1660–70.

18. Lee W, Chan H, Chan P, Fiorillo T, Fiveland E, Foo T, et al. A Magnetic Resonance Compatible E4D Ultrasound Probe for Motion Management of Radiation Therapy. IEEE Netw. 2017 Sep;2017:10.1109/ULTSYM.2017.8092223.